\documentstyle[eqsecnum,aps,psfig]{revtex}
\begin{document}
\draft
\preprint{}
\title{Astronomical Constraints on the Cosmic Evolution of \\
the  Fine Structure Constant and Possible Quantum 
Dimensions\footnote{Phys Rev Lett. 2001, 85, 5511}}
\author{C.L. Carilli$^1$, K.M. Menten$^2$, J.T.Stocke$^3$, 
E. Perlman$^4$,R. Vermeulen$^5$, \\
F. Briggs$^6$,  A.G. de Bruyn$^{5,6}$, J.Conway$^7$,C.P.Moore$^{8}$
}
\address{
$^1$National Radio Astronomy Observatory, Socorro, NM, USA 87801\\
$^2$Max-Planck-Institute f\"ur Radioastronomie, Auf dem H\"ugel 69, 
D-53121, Bonn, Germany \\
$^3$University of Colorado, Boulder, CO, USA\\
$^4$Space Telescope Science Institute, Baltimore MA\\
$^5$Netherlands Foundation for Research in Astronomy, Dwingeloo,
the Netherlands\\
$^6$Kapteyn Research Institute, Groningen, the Netherlands\\
$^7$Onsala Space Observatory, Onsala, Sweden\\
$^8$Smithsonian Astrophysical Observatory, 60 Garden St., Cambridge,
MA, 02138, USA \\
}
\date{\today}
\maketitle
\begin{abstract}
We present measurements of absorption by the 21cm hyperfine transition 
of neutral hydrogen toward radio sources at substantial
look-back times. These data are used in combination with
observations of rotational transitions of
common interstellar molecules, to set 
limits on the evolution of the fine structure constant:
${{\dot\alpha}\over{\alpha}} < 3.5 \times 10^{-15}$~year$^{-1}$,
to a look-back time of 4.8 Gyr.
The neutral hydrogen observations employed Very Long Baseline
Interferometry in order to mitigate the substantial
uncertainty arising from the fact that observations at
very different wavelengths may probe different lines-of-site due to 
frequency dependent structure of the background source. 
We discuss the implication of these results on theories
unifying natural forces based on  compact quantum
dimensions. In the context of string theory, the 
limit on the secular evolution of 
the scale factor of these compact dimensions, $R$, is 
${{\dot R}\over{R}} < 10^{-15}$~year$^{-1}$. Including
terrestrial and other astronomical measurements 
places limits (2$\sigma$) on  slow oscillations of   $R$ 
from the present to the epoch of cosmic nucleo-synthesis, just 
seconds after the big bang, of 
${{\Delta R}\over{R}} < 10^{-5}$.
\end{abstract}


\widetext 

Since the startling
realization that we live in an evolving universe, there have 
been numerous hypotheses concerning the cosmic evolution of physical
constants.
Dirac\cite{dirac} pointed out that the ratio between the strength
of the gravitational force to that of the electromagnetic 
force  $\approx 10 ^{39}$
$\approx$  the age of the universe measured in atomic
units of time.
(ie. 'le tempon' = light crossing time of the
classical radius of the electron). 
He speculated that such large dimensionless numbers
may be more fundamental than the supposed fundamental constants, 
and that the constants evolve in such a way that observers
at  any given epoch reach similar conclusions.
This line of reasoning leads to a variation of the
gravitational constant\footnote{Study of the variation of
fundamental constants must consider dimensionless forms
involving products of the constant in question with $\hbar$, $m_p$, 
and $c$, since dimensionless quantities are
invariant under coordinate transformation, for example:
$\gamma \equiv {{G m_p^2}\over{\hbar c}}$ or 
$\alpha \equiv {{e^2}\over{\hbar c}}$\cite{dyson,barrow}.
We use $\rm H_o = 75~ km~ s^{-1}$
Mpc$^{-1}$, and a deceleration parameter, q$_o$ = 0.5.}, $G$,
over cosmic time of order: 
${{\dot G}\over{G}} \sim \rm H_o \sim 10^{-10}$ year$^{-1}$,
where $\rm H_o$ is the Hubble constant quantifying the local 
expansion of the universe.
A similar prediction on $G$ comes from
the Brans-Dicke equation relating mass/energy to a cosmic scalar
field \cite{weinberg}.
Variation of $G$ at the Hubble-rate
has been ruled out by measurements of planetary orbits \cite{dyson,varpot}.  
Gamow\cite{gamow} speculated that the charge on the
electron, $e$, or more precisely, 
the fine structure constant, $\alpha$,
varies and not $G$. This would lead to variation
of the Rydberg constant 
and hence contribute to the  observed redshift, $z$, of spectral
features in distant objects normally attributed to the evolution
of the 4-dimensional (4D) cosmic scale factor. 
Lastly, Sisterna and Vucetich\cite{sisterna} point out that 
stringent limits to the variation of physical constants support
Einstein's strong equivalence principle, and hence provide
`accurate verification' for general relativity as the correct
low-energy theory of gravity. 

The idea of cosmic variation of physical constants has been revisited
with the advent of models unifying the forces of
nature based on the symmetry properties of 
quantum dimensions, such as the Kaluza-Klein (KK) model,
or the more general requirement of extra dimensions in 
superstring theory (SS) \cite{schwartzseiberg,kolbturner}. 
These extra dimensions
have a scale factor, $R$, of order the Planck scale, $R \approx
10^{-33}$ cm, 
and manifest themselves only during the first instant of 
creation, corresponding to the Planck time, $10^{-43}$ seconds
after the big bang, or at energies above $10^{19}$ GeV. 
These compact 
dimensions quickly vanish during the cosmic expansion of
our familiar 4D space-time, however they may
still have observable consequences, since 
the constants of nature observed in 4D are the
result of integration over the extra dimensions\cite{kolbturner}.
It has been
hypothesized that a variation of $R$ with cosmic epoch
could lead to a variation of the physical constants measured in
4D\cite{barrow,marciano}.   
In particular, the fine structure constant, $\alpha$,
is predicted to behave as $R^{-2}$ in 
KK theories, and $R^{-6}$ in SS theories\cite{barrow,kolb}.
Unfortunately, while the time dependence 
of the cosmic scale  factor in 4D space-time is well understood in
terms of Einstein's theory of gravity, there is no analogous prediction
concerning the evolution of
$R$ in extra-dimensional theories. It is possible
that $R$  increases or decreases  monotonically, or even
oscillates,  with cosmic time\cite{marciano}. 
Still, a measurement of the cosmic evolution of $\alpha$ would 
provide qualitative supporting evidence for the existence of 
compact dimensions\cite{levshakov}.


Limits to the evolution of $\alpha$ include laboratory
measurements\cite{prestage}, consideration of the abundances of
radioactive isotopes\cite{prestage}, and consideration of fluctuations
in the microwave 
background and other cosmological constraints \cite{avelino,livio}. 
The most stringent terrestrial
measurement comes from the Oklo natural reactor, which occurred
about 1.8 Gyr ago. The 95$\%$ confidence limit (2$\sigma$) from these 
calculations is $|{{\dot\alpha}\over{\alpha}}| < 
7 \times 10^{-17}$~year$^{-1}$\cite{damour}. 
Calculations of primordial nucleo-synthesis allow for a maximum
variation of $|{{\dot\alpha}\over{\alpha}}| < 
11 \times 10^{-15}$~year$^{-1}($\cite{kolb,barrow}
although see \cite{bergstrom}). 
Accurate spectroscopy of absorption and emission 
lines from objects at cosmologically significant redshifts 
can be used to set limits on the evolution of physical
constants\cite{savedoff,bahcall,wolfe,levshakov,varpot}, 
Such  studies lead to limits of\cite{webb,cowie}:
$|{{\dot\alpha}\over{\alpha}}| < 2 \times 10^{-15}$~year$^{-1}$.
 
Drinkwater et al.\cite{drinkwater} and Wiklind and
Combes\cite{wiklind} have compared absorption by molecular rotational
transitions at millimeter wavelengths to 
HI 21cm absorption to determine the evolution of the product $Y \equiv
g_p \alpha^2$, where $g_p$ is the proton-to-electron magnetic moment.
These cm and mm measurements have the potential 
advantage over optical spectroscopy in that spectral
resolutions of 1 km s$^{-1}$ or better are easily obtained, and the
absorption lines themselves can be extremely narrow\cite{wiklind},
with the ultimate limitation  ($\approx 10^{-8}$) 
being the accuracy of the 
lab measurements of the transitions in question.
Based on errors due to signal-to-noise alone, the current 
best limit (2$\sigma$) using this technique is 
$|{{\dot\alpha}\over{\alpha}}| < 1 \times 10^{-15}$~year$^{-1}$.
However, these measurements have  a potentially much larger
uncertainty arising from possible differences in 
the velocities of the HI 21cm and 
molecular absorbing gas within a given
galaxy\cite{wiklind,levshakov,cowie}. Velocity differences can arise 
both along a given line-of-site, and also due to the fact that
observations at very different wavelengths (eg. mm versus cm) 
may probe different lines-of-site due to 
frequency dependent spatial structure of the background source. 
If line-of-sight differences occur on kilo-parsec (kpc)
scales, then systematic velocity differences can arise due to the galaxian
potential, and can be of order 100 km s$^{-1}$. If line-on-site
differences can be limited to sub-kpc scales, then it can be argued
that the residual uncertainty is likely to be of order 10
km s$^{-1}$, ie. comparable to the typical velocity dispersion
of the interstellar medium in galaxies \cite{vanderkruit}.


In this paper we present observations of HI 21cm 
absorption at intermediate redshifts using the technique of Very Long
Baseline Interferometry (VLBI)\cite{thompson}. These observations
provide spatial resolutions of tens of milli-arcseconds (mas), 
corresponding to sub-kpc spatial scales, and hence mitigate the
potential problem of probing different lines-of-sight at different
wavelengths. 
In the following analysis we will quote heliocentric
redshifts, $z$, and differences in redshifts, $\Delta z$.
In terms of observed frequency differences, $\Delta \nu_o$, 
source-frame velocity differences, $\Delta v$, and
the variation of the physical quantity in question, $\Delta Y$:
${{\Delta Y}\over{Y}} = 
{{\Delta z}\over{1 + z}} = 
{{\Delta \nu_o}\over{\nu_o}} \sim
{{\Delta v}\over{c}}$. 

We have observed the HI 21cm absorption toward the cosmic radio sources
0218+357 and 1413+135 using VLBI techniques (Fig. 1). The source
0218+357  is a gravitationally lensed background radio source, 
and the absorption is by gas in  the lensing galaxy\cite{carilli1,wiklind2}.
The source 1413+135 is a radio
loud AGN at the center of an edge-on spiral galaxy, and the absorption is
by gas in a molecular cloud in
the disk of the parent galaxy\cite{carilli2,wiklind}. 
The data were processed at the VLBA correlator in Socorro NM,
and images were generated
using the Astronomical Image Processing System (AIPS).
The data were fringe fit starting with a point source model for
1413+135 and a double source for 0218+357,
and then imaged using standard hybrid imaging
techniques\cite{thompson}.  
In each case spectral data image 
cubes of 256 channels over 4 MHz bandwidth were analyzed. 
The details of the observations and their astrophysical implications
will be presented elsewhere. Herein we concentrate on the information
relevant to constraining the evolution of $\alpha$. 

For 0218+357 the continuum image has a resolution of 80 mas, 
corresponding to a physical scale of 390 pc, and
shows two lensed radio components separated by 350mas. We find that
the dominant HI 21cm absorption line is toward the southwest 
component. This is also true for the molecular
absorption\cite{menten}. We have fit a two component 
Gaussian model to the absorption spectrum using a least squares
algorithm.  The principal component has a peak optical depth,
$\tau = 0.12$, and a Full Width at Half Maximum, FWHM = 
15 km s$^{-1}$, at a redshift of 
$z = 0.684676 \pm 0.000005$. 
The error corresponds to a change in $z$ 
leading to a unit change in reduced $\chi^2$.
The peak of the molecular absorption in this system is at
$z = 0.684693 \pm 0.000001$\cite{wiklind}, hence the difference
between the molecular and HI 21cm redshifts is: ${{\Delta z}\over{1 +
z}} = 1.0\times 10^{-5} \pm 3\times 10^{-6}$.

For 1413+135 the continuum image has a resolution of 20 mas,
corresponding to a spatial scale of 70 pc, and shows 
a core-jet morphology extending
over 60 mas. The position of the inverted-spectrum
nucleus is shown as a large cross, and again, at mm wavelengths
only the
nuclear radio component is detected\cite{perlman}.
At 1143 MHz the jet  dominates the total radio continuum
emission from 1413+135. The absorption spectrum at the peak
surface brightness of the jet, indicated by a small cross,
is shown in Figure 1. The sensitivity of these observations 
is insufficient to detect absorption toward the 
weaker nuclear radio component. 
We fit the spectrum of the jet  with a one component 
Gaussian model, resulting in
$\tau = 0.89$ and  FWHM = 15 km s$^{-1}$, at  
$z = 0.246693 \pm 0.000001$. 
The molecular 
absorption in this system is extremely narrow (FWHM $\approx 1$ km
s$^{-1}$) with a peak at  $z = 0.2467091 \pm
0.0000003$\cite{wiklind}, hence the difference
between the molecular and HI 21cm redshifts is: ${{\Delta z}\over{1 +
z}} = 1.3\times 10^{-5} \pm 8\times 10^{-7}$.


The errors given above are 1$\sigma$ strictly based on the
fitting process and the  signal-to-noise ratio (SNR)
in the observed spectra. There are a number of other systematic errors 
in the measurement and Gaussian fitting process which must be
considered, including: (i)
the transfer of the sky frequencies from the
telescopes through the correlator to the data reduction programs,
and (ii) the interpolation process when deriving 
redshifts to accuracies better than the spectral 
channel width of 15~kHz, including 
possible variations in the zero intensity level, 
the number of Gaussians
used in the modeling, and the assumption of a
Gaussian shape for the line profile. The results 
indicate that the measurement errors other than SNR 
are $<$ 10 kHz, corresponding to  3 km s$^{-1}$,
or ${{\Delta z}\over{1 + z}} < 1\times10^{-5}$. 

The sum total of the uncertainties discussed above are comparable to
the most accurate astronomical measurement to date\cite{webb,cowie}, 
and could be improved with higher SNR, higher spectral resolution 
observations. However, there remains the 
possibility of systematic velocity offsets between the HI  21cm
absorbing gas and the molecular absorbing gas in a given galaxy.
Drinkwater et al.\cite{drinkwater} demonstrate that
the velocity of HI 21cm absorption in the Galaxy shows a statistical
correlation with that of molecular absorption, with a dispersion of
only 1.2 km s$^{-1}$. On the other hand, there are
clearly Galactic and extragalactic examples in which significant
velocity differences occur\cite{wiklind}, and a 
statistical analysis of the type in Ref.\onlinecite{drinkwater} 
will be rigorously valid only when applied 
to a statistical sample of absorbers. Note that this error does 
not apply to calculations of ${{\dot\alpha}\over{\alpha}}$
based on fine structure absorption 
doublets by alkaline metals\cite{webb}.

For 1413+135 the molecular and HI 21cm absorption spectra are
narrow in velocity, 
suggesting absorption by the general ISM of the host galaxy.
But the HI 21cm absorption occurs along a line-of-site
toward the jet which is offset from that toward the nucleus by
25mas.  This line-of-site offset will give rise to a velocity 
difference less than 10 km s$^{-1}$ as long as the 
absorbing gas is beyond 0.5 kpc from the nucleus, 
assuming a flat rotation curve for the galaxy with a velocity of 200 km
s$^{-1}$. For 0218+357 the absorption is again narrow,
and our VLBI observations probe a region of 
diameter $< 390$ pc through
a face-on spiral galaxy \cite{carilli1}. 
In summary, it is reasonable to assume that the 
dominant uncertainty between 
the molecular and HI 21cm redshifts in both 1413+135 and
0218+357 is due to small-scale (sub-kpc) ISM motions, 
i.e. $\approx  10$ km s$^{-1}$. 
A velocity  uncertainty of  10 km s$^{-1}$ 
leads to a limit to the evolution of
the fine structure constant of 
$|{{\dot\alpha}\over{\alpha}}| < 3.5 \times 10^{-15}$~year$^{-1}$
to a look-back time, $t$, of 4.8 Gyr  for 0218+357, 
and $|{{\dot\alpha}\over{\alpha}}| < 6.7 \times 10^{-15}$~year$^{-1}$
to $t = 2.5$ Gyr 
for 1413+135, assuming that $g_p$ is constant\cite{tubbs}. 
These limits are sufficient to rule-out hypotheses 
D and E summarized in Dyson\cite{dyson}, in which 
$\alpha \propto t$, and $\alpha \propto \rm log(\sl t)$, 
respectively.

In Table 1 we summarize the limits to
${{\Delta \alpha}\over{\alpha}}$ over time, and to 
${{\dot R}\over{R}}$ in KK and SS theories.
The most stringent  limit to secular evolution 
of $R$ remains the Oklo limit, although
this limit has been called into question recently \cite{sisterna}. 
Hence, limits using other methods
provide an important check of the Oklo limit. Also,  
all the limits taken together argue strongly against a slowly
oscillating $R$ with ${{\Delta R}\over{R}} < 10^{-5}$  over the entire
history of the universe. And  the limits to
${{\dot\alpha}\over{\alpha}}$  
all have different functional dependencies on $\alpha$ and other
physical constants.  
The lack of variation observed for any of the different
products of physical constants argues against models of 
a `cosmic conspiracy' in which the individual constants vary 
in concert to result in a given observable remaining invariant. 
One method for overcoming the uncertainty due to small scale ISM 
motions is to
observe a large sample of sources ($> 100$), and rely on the statistical
correlation described in Drinkwater et al.\cite{drinkwater}.
Study of such a large
sample will only become possible with the greatly increased
sensitivity of the Atacama Large Millmeter Array, and the
Square Kilometer Array.

\acknowledgments

The National Radio Astronomy Observatory is operated by Associated
Universities Inc.. under cooperative agreement with the National
Science Foundation. CLC would like to acknowledge partial support from 
from the Alexander von Humboldt Society.
CLC and FB would like to thank R. Kolb and F. Dyson 
for useful discussions concerning this work.

\clearpage
\newpage
                         
\begin{table}
\caption{Limits (2 $\sigma$) to the evolution of the fine structure constant
and the scale factor of compact dimensions. Quantities with redshifts
are based on astronomical spectroscopy.
\label{table1}}
\begin{tabular}{ccccccc}
Method & $z$ & Look-back Time & ${\Delta \alpha}\over{\alpha}$ & 
${{\dot R}\over{R}}$ (SS) & ${{\dot R}\over{R}}$ (KK) &  Reference \\
~ & ~ & Gyr & $\times 10^{-6}$ & $\times 10^{-16}$
year$^{-1}$ & $\times 10^{-16}$ year$^{-1}$ & ~ \\
\tableline

Laboratory & -- & 10$^{-9}$ & $4 \times 10^{-8}$ & 70  & 200  & 
\onlinecite{prestage} \\

Oklo Reactor & -- & 1.8 & 0.13  & 0.12 & 0.35  & \onlinecite{damour}\\

1413+135 & 0.25 & 2.5 & 17 & 11 & 33 &  This paper\\ 

0218+357 & 0.68 & 4.8 & 17 & 6  & 18 &  This paper\\ 

Radioactive elements & -- & 5 & 24 & 8 & 25 &  \onlinecite{kolb} \\

Fine Structure & 0.8 & 5.2 & 8 & 2.6 & 7.7 &  \onlinecite{webb} \\

Fine Structure & 1.3 & 6.3 & 8 & 2.1 & 6.3 &  \onlinecite{webb} \\

Metal + HI 21cm & 1.78 & 7.0 & 15 & 3.5 & 10 & \onlinecite{cowie} \\ 

Primordial Nucleosynthesis & -- & 8.9 & 100 & 18 & 55 & \cite{kolb} \\
\end{tabular}
\end{table}

\begin{figure}
\caption{The contour images (left) show total radio continuum surface
brightness distributions for 0218+357 (top) and 1413+135 (bottom)
derived from VLBI observations at
the frequency of the redshifted HI 21cm absorption lines. 
The observations of 0218+357 were made using the European VLBI Network 
in October 1998, while those for 1413+135 were made using the
Very Long Baseline Array operated by the
National Radio Astronomy Observatory during July 1998.
The spatial resolutions are shown in the insets (FWHM).
The total flux density of 1413+135 at this frequency is 1.2 Jy
while that for 0218+357 is 1.8 Jy.
The contour levels are logarithmic in the square root two with
an arbitrary absolute scale since no {\sl a priori} gain calibration was 
applied due to lack   of accurate system temperature measurements
at these low frequencies. 
The spectra (right) show optical depth in redshifted 
HI 21cm absorption toward the positions designated by the crosses in
the continuum images. The solid line is the measured data, and the
long dash line is the Gaussian model fit. 
In the case of 1413+135,
the position of the AGN is shown with the large cross, but the
HI 21cm absorption spectrum was made at the position of the 
peak surface brightness of the jet, corresponding to the smaller 
cross. The frequency scale has been corrected to a heliocentric
rest-frame. The short dash lines show the 
CO absorption line profiles, scaled in frequency accordingly and
on an arbitrary optical depth scale, 
as observed by the IRAM 30m telescope
(see [25] and references therein).
The arrows indicate the expected HI frequencies based on Gaussian 
model fits to the molecular absorption lines. For 0218+357 the profile 
displayed is for CO(1-2), while that for 1413+135 is CO(0-1).
}
\label{autonum}
\end{figure}

\begin{figure}
\vspace*{-1in}
\psfig{figure=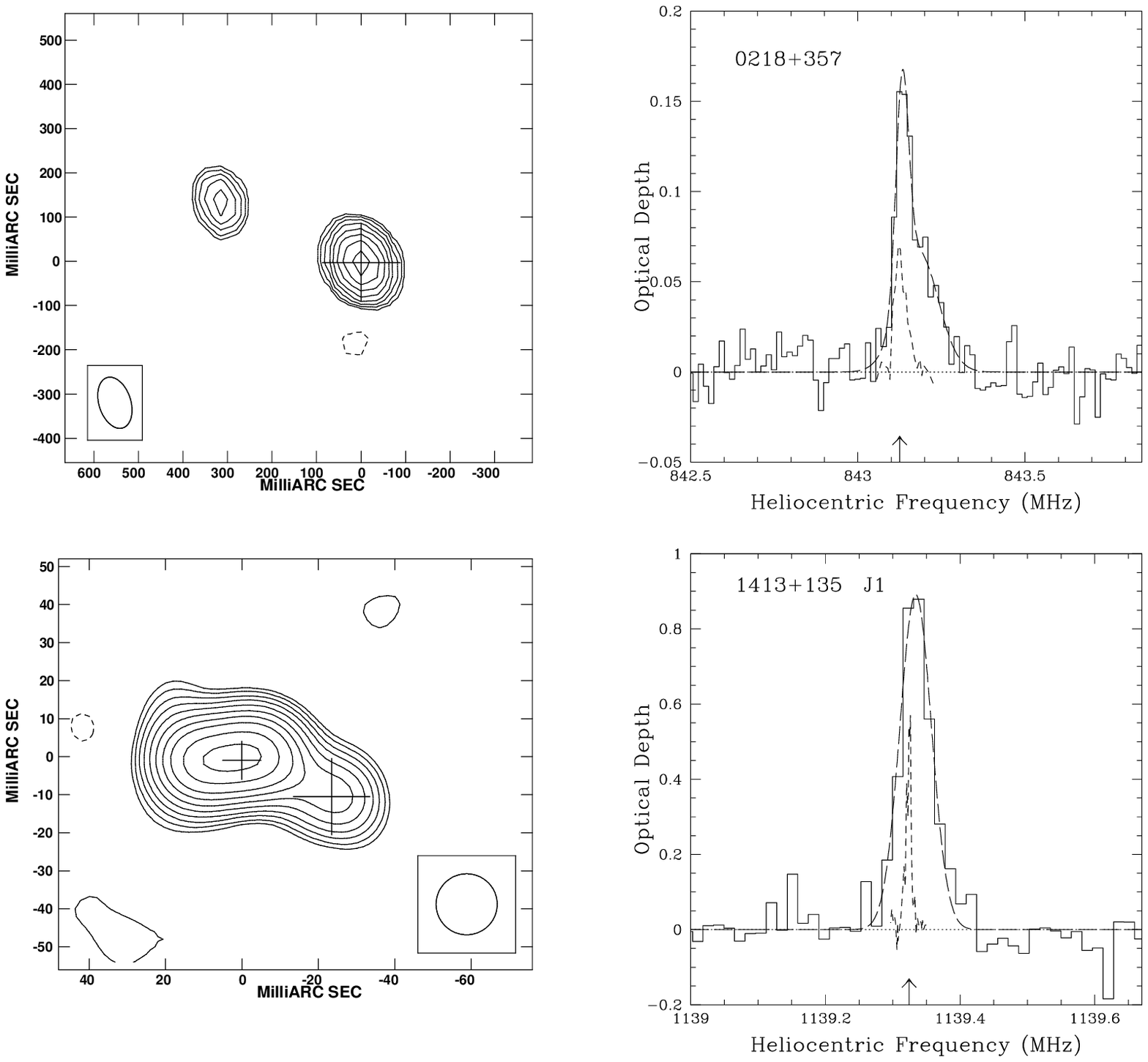,width=8in}
\end{figure}

\end{document}